# Magnetic properties of undoped Cu$_2$O fine powders with magnetic impurities and/or cation vacancies

Chinping Chen[1a], Lin He[1], Lin Lai[1], Hua Zhang[2], Jing Lu[1], Lin Guo[2b], and Yadong Li[3]

[1]Department of Physics, Peking University, Beijing, 100871, People's Republic of China

[2]School of Chemistry and Environment, Beijing University of Aeronautics and Astronautics, Beijing, 100083, People's Republic of China

[3]Department of Chemistry, Tsinghua University, Beijing, 100084, People's Republic of China

Abstract:

Fine powders of micron- and submicron-sized particles of undoped Cu$_2$O semiconductor, with three different sizes and morphologies have been synthesized by different chemical processes. These samples include nanospheres 200 nm in diameter, octahedra of size 1 $\mu$m, and polyhedra of size 800 nm. They exhibit a wide spectrum of magnetic properties. At low temperature, $T$ = 5 K, the octahedron sample is diamagnetic with the magnetic susceptibility, $\chi_{OH}$ = -9.5×10$^{-6}$ emu.g$^{-1}$.Oe$^{-1}$. The nanosphere is paramagnetic with $\chi_{NS}$ = 2.2×10$^{-5}$ emu.g$^{-1}$.Oe$^{-1}$. The other two polyhedron samples synthesized in different runs by the same process are found to show different magnetic properties. One of them exhibits weak ferromagnetism with $T_C$ ~ 455 K and saturation magnetization, $M_S$

---

[a]Corresponding : cpchen@pku.edu.cn, Phone : +86-10-62751751
[b]guolin@buaa.edu.cn

~ 0.19 emu/g at $T$ = 5 K, while the other is paramagnetic with $\chi$ = $1.0\times10^{-5}$ emu.g$^{-1}$.Oe$^{-1}$. The total magnetic moment estimated from the detected impurity concentration of Fe, Co, and Ni, is too small to account for the observed magnetism by one to two orders of magnitude. Calculations by the density functional theory (DFT) reveal that cation vacancies in the Cu$_2$O lattice are one of the possible causes of induced magnetic moments. The results further predict that the defect-induced magnetic moments favour a ferromagnetically coupled ground state if the local concentration of cation vacancies, $n_C$, exceeds 12.5%. This offers a possible scenario to explain the observed magnetic properties. The limitations of the investigations in the present work, in particular in the theoretical calculations, are discussed and possible areas for further study are suggested.

# 1. Introduction

The *p*-type semiconducting $Cu_2O$ is an oxide semiconductor with a direct band gap of about 2.0 eV. With a large hole mobility at room temperature, ~100 $cm^2V^{-1}s^{-1}$, $Cu_2O$ is very promising for applications in optoelectronics and as a diluted magnetic semiconductor (DMS) [1]. The magnetic properties of this material with and without magnetic doping are therefore of fundamental interest to the spin polarized transport behaviours. For the magnetic properties of doped samples, observations of room temperature ferromagnetism in the Al and Co co-doped $Cu_2O$ film [2] and in the Mn doped $Cu_2O$ films [3,4] have been reported. In the early days, Zener proposed that the indirect exchange interaction between localized *d*-shell electrons via the itinerant conduction electrons leads to ferromagnetic (FM) coupling [5]. By this theory, the ferromagnetism of Fe, Co, and Ni has been successfully accounted for [6]. Later on, the theory for the indirect exchange coupling via the itinerant electron was further developed by Ruderman and Kittel [7], Kasuya [8], and Yosida [9] (RKKY). For DMS systems, the RKKY interaction was proposed to describe the exchange coupling between the doped magnetic impurities [10-12]. On the other hand, with pure $Cu_2O$ where neither $Cu^{+1}$ nor $O^{2-}$ is magnetic ion and the *d* shell of $Cu^{+1}$ is full, the magnetic properties are diamagnetic [13]. In experiments, pure $Cu_2O$ is found to show diamagnetic or paramagnetic (PM) properties, with the room temperature susceptibility of $\chi$ = -0.2×$10^{-6}$ [14], 0.3×$10^{-6}$ [15] and -0.155×$10^{-6}$ emu/g-Oe [16]. In addition, PM moments in the diamagnetic background have also been observed [16]. It is believed that the detected magnetic moments arise from the presence of cation vacancies which are due to the oxygen excess during the treatment of the $Cu_2O$ sample [16]. This suggests a process-dependent magnetic property with an undoped $Cu_2O$ sample.

Recently, weak ferromagnetism has been reported with nonmagnetic $HfO_2$ film [17]. It has been ascribed to the presence of Hf vacancies by the DFT calculation [18]. An *ab initio* calculation for the electronic structure shows that Hf vacancies will lead to ferromagnetically coupled high-spin defect states. This is the case for CaO in which $Ca^{2+}$ vacancy is predicted to exhibit local magnetic moment by an *Ab initio* calculation [19]. Weak ferromagnetism has also been observed with nonmagnetic $TiO_2$ and $In_2O_3$ films [20,21]. It is also attributed to the defect states. Sundaresan *et al.*, even claimed that ferromagnetism is a universal property of nonmagnetic oxides nanoparticles [22]. They have observed room temperature ferromagnetism in nonmagnetic oxides such as $CeO_2$, $Al_2O_3$, ZnO, $In_2O_3$, and $SnO_2$, with diameters ranging from 7 to 30 nm. The origin of the ferromagnetism has been attributed to the oxygen defects [22]. It is revealed by the above-mentioned experiments and theoretical calculations that defect vacancies are an important cause for the presence of weak magnetism with a nonmagnetic material. However, there are also different opinions. For example, Osorio-Guillén *et al* [23] have pointed out that it is impossible to achieve sufficient vacancies needed in CaO to account for the ferromagnetism even with the most favorable growth conditions at equilibrium. This is attributed to the extremely high formation energy of a vacancy. Experimentally, Abraham *et al* [24] and Rao *et al* [25] did not observe ferromagnetism with undoped $HfO_2$ films. More interestingly, a measurable FM signal with a $HfO_2$ film is reported and is ascribed to the contamination from the stainless-steel tweezers [24]. In the work by Sundaresan *et al* [22], the observed saturation magnetization for a series of nonmagnetic oxides are on the order of $10^{-4} \sim 10^{-3}$ emu/g. An impurity concentration of $5\times10^{-7} \sim 5\times10^{-6}$ g-Fe/g-sample (0.5 ~ 5 ppm) is sufficient to explain the observed ferromagnetism. This is far beyond the detection sensitivity of the instruments used in most of the reported experiments [17,20-23], for example, x-ray diffraction (XRD) with a sensitivity of a few percent and energy-dispersive x-ray (EDX) analysis with a sensitivity of a few parts in a thousand.

Nanoparticles of $Cu_2O$ with different sizes and morphologies have been synthesized [26-31]. As an example, nano-cubes with average edge length varying from 200 to 450 nm have been reported [26]. In this article, we report investigations on the magnetic properties of micron- and submicron-sized, undoped $Cu_2O$ semiconductors by magnetization measurements and study the possible origin of the observed magnetism by DFT calculations. For the experiment, it is interesting to observe a wide spectrum of magnetic properties, from diamagnetic to ferromagnetic, with a series of pure $Cu_2O$ particles synthesized by different processes. The concentration of magnetic impurities, such as the Fe, Co, and Ni, in the samples has been analysed by EDX and inductively coupled plasma-atomic emission spectroscopy (ICP-AES). The impurity concentration with these samples is beyond the sensitivity of XRD and EDX. According to the detection by ICP-AES, the concentration of the magnetic impurities contained in the samples is not high enough to explain the observed magnetism. The DFT calculation, on the other hand, has revealed that a good part of the magnetism observed with the undoped $Cu_2O$ samples in the present experiments, which is too large to be accounted for by the detected magnetic impurities alone, may be attributed to the presence of cation vacancies. In addition, the DFT calculations for the ground state energy of the cation-vacancy-induced magnetic moments reveal that it is possible for these induced moments to exhibit both PM and FM properties, depending on the local concentration of cation vacancies. Particularly, a ferromagnetically coupled ground state is predicted with a local concentration of cation vacancies, $n_C$, exceeding 12.5%. More interestingly, the total magnetic moments corresponding to $n_C = 12.5\%$ along with the contribution from the detected impurities almost account for the observed magnetism.

## 2. Sample preparation and characterization

The magnetic properties of pure $Cu_2O$ are investigated with four powder samples synthesized by three different chemical methods. The experimental details for two of them, which are nanospheres of 200 nm in diameter and octahedra of 1 μm in size, have been reported in a previous paper [32]. The experimental details for the remaining two, which are both polyhedra of about 800 nm in size, are presented in this work. The powder samples of pure submicron-sized, polyhedron $Cu_2O$ particles have been synthesized chemically via a solution-phase route with the assistance of a polymer reagent. A typical process involves several steps. First, about 1.10 g of poly(vinylpyrrolidone) (PVP; molecule weight 30 000) and 0.17 g of $CuCl_2·2H_2O$ were dissolved in 100 ml of distilled water, which was stirred with a magnetic stirrer. The solution was then heated in a water bath at a constant temperature of 54 °C, and stirred for 30 min to ensure that the PVP and $CuCl_2 \cdot 2H_2O$ dissolved completely. The solution appeared to be lucid light green. Afterwards, 10 ml of 2 M NaOH was added drop by drop into the above solution under constant stirring. The solution turned into turbid blue green, then dark brown, because of the formation of precipitates in the solution. After stirring for another 20 min, 10 ml of 0.6 M ascorbic acid solution was added drop by drop into the dark brown solution, with constant stirring. Then, a turbid red liquid, which was opaque due to the appearance of precipitates, gradually formed. The solution was heated continuously in a water bath at 54 $^o$C for 3 h. Then, a red $Cu_2O$ precipitate was produced. Centrifugation was carried out to separate the precipitates, which were then rinsed with distilled water several times and subsequently with absolute ethanol to remove the polymer completely. Finally, the volatile solvent was evaporated in vacuum at 80 °C, yielding a loose red powder.

The structural and compositional characterizations were performed by powder XRD using a Rigaku D\Max-2200, by scanning electron microscopy (SEM) at an accelerating voltage of 15 kV using a FEI Siron 200, by transmission electron microscopy (TEM) on a JEM-2100F using an accelerating voltage of 200 kV, by high-resolution TEM (HRTEM) and EDX using a Hitachi S-4300, and by ICP-AES using a VISTA-MPX. The magnetic measurements were carried out on a Quantum Design MPMS SQUID magnetometer.

Figure 1(a) shows the XRD pattern for the polyhedron sample in the log-$Y$ scale using Cu $K_\alpha$ radiation. It shows a high degree of crystallinity. All of the Bragg peaks are indexed to the pure cubic phase of $Cu_2O$ [33]. The (200) peak has a higher intensity compared with the standard powder pattern. It indicates a growth of preferential orientation. There is no other Bragg peak detected arising from any possible impurity phase, such as Cu, CuO or any magnetic impurity such as Fe, Ni, or Co. The lattice constant of the pure $Cu_2O$ unit cell derived from the XRD analysis is 0.4267 nm.

Figures 1(b) and 1(c) present typical SEM images of the $Cu_2O$ polyhedra at different magnifications, 5000 times for figure 1(b), and 20000 times for figure 1(c). The $Cu_2O$ crystals exhibit uniform and regular polyhedron morphology with a smooth surface. The average main edge length of the polyhedra is estimated as about 800 nm. More detailed information on the as-prepared $Cu_2O$ samples has been acquired by the HRTEM investigation. Figure 1(d) shows the HRTEM image of the $Cu_2O$ polyhedra. The lattice fringes are clearly visible with the two lattice planes, (111) and ($1\bar{1}1$), marked in the figure. A higher magnification image for the same region is shown in the inset. The interplanar distance of 0.246 nm matches well with the value derived from the XRD (111) peak, and the angle, 70.53º, between the two lattice planes, (111) and ($1\bar{1}1$), coincides with the calculated value, 70.48º, for the cubic lattice structure.

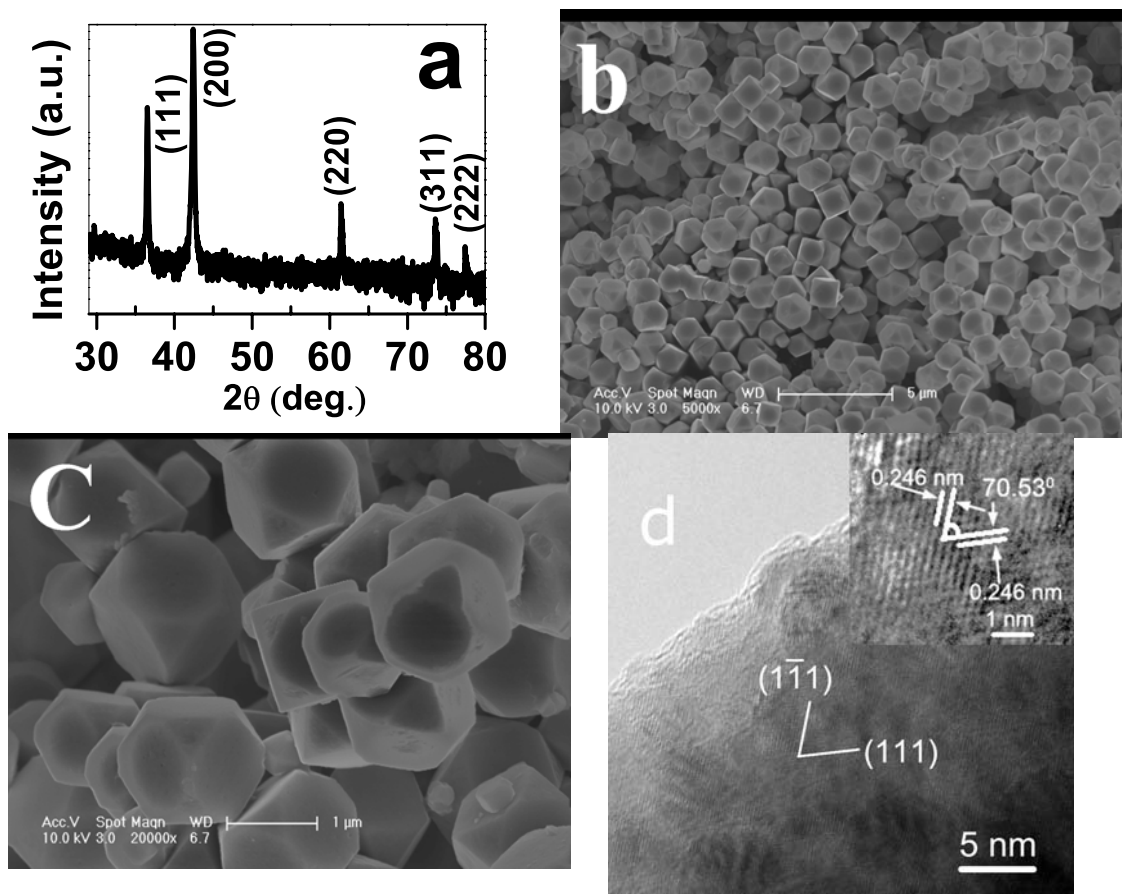

Figure 1. **(a)** Powder XRD pattern of the polyhedron $Cu_2O$. The *Y* axis is in the log scale to show a better view in the low intensity region. **(b)** SEM image of the polyhedron $Cu_2O$. The scale bar is 5 μm. **(c)** SEM image at a higher magnification. The scale bar is 1 μm. **(d)** HRTEM image of the as-prepared $Cu_2O$. The marked

region shows the lattice fringes of (111) and ($1\bar{1}1$) planes. A higher magnification image is displayed in the inset. The interplanar distance is determined as 0.246 nm.

## 3. Magnetic measurements and analysis

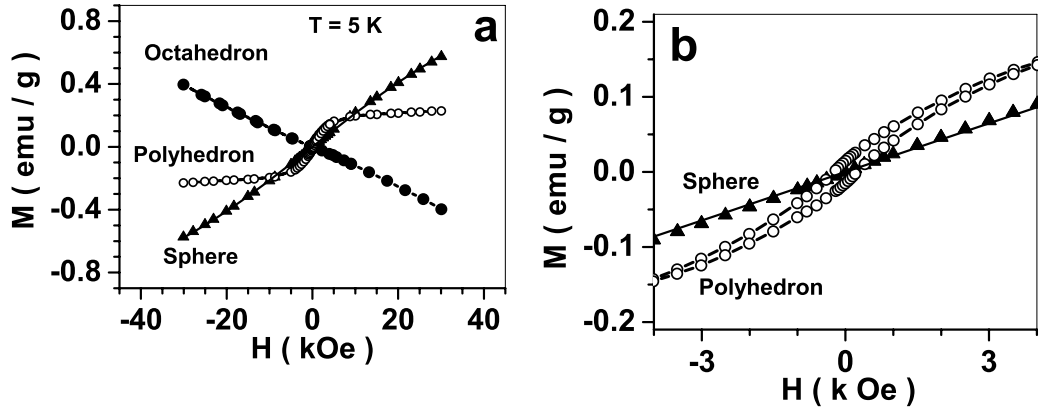

Figure 2. **(a)** $M(H)$ curves measured at $T = 5$ K for the $Cu_2O$ samples of diamagnetic octahedra, PM nanospheres, and FM polyhedra. **(b)** Low field region of the $M(H)$ curves shown in (a) for the PM nanospheres and the FM polyhedra.

The magnetic measurements were performed for the powder samples, which are 5.34 mg of diamagnetic octahedra, 18.49 mg of FM polyherdra, 1.11 mg of PM polyhedra and 12.97 mg of PM nanospheres, by the zero-field-cooling (ZFC) and field-cooling (FC) $M(T)$ measurements and the field dependent $M(H)$ measurements. The field dependent magnetization, $M(H)$, measurements at $T = 5$ K for three of the samples are shown in figure 2(a). Data for the PM polyhedra are not shown. The magnitude of magnetization shown in the figure is without the correction of background (BKGD) diamagnetism for the capsules holding the samples in the measurements. The diamagnetic susceptibility for a typical capsule used in the experiment is measured as, $\chi_{capsule} \sim -4.80 \times 10^{-7}$ emu.g$^{-1}$.Oe$^{-1}$ at 5 K (-4.54 $\times 10^{-7}$ emu.g$^{-1}$.Oe$^{-1}$ at 300 K). This value is almost temperature-independent, within 5% from 5 to 300 K. The magnitude of correction for the BKGD diamagnetic signal is then calculated according to $\Delta\chi_{BKGD} \sim \chi_{capsule}(m_{capsule}/m_{sample})$. The susceptibility of the diamagnetic octahedron sample is then determined as $\chi_{OH}$ = -9.5×10$^{-6}$ emu.g$^{-1}$.Oe$^{-1}$ (already corrected for $\Delta\chi_{BKGD}$ ~ -3.5×10$^{-6}$ emu.g$^{-1}$.Oe$^{-1}$), the PM nanosphere sample, $\chi_{NS}$ = 2.2×10$^{-5}$ emu.g$^{-1}$.Oe$^{-1}$ ($\Delta\chi_{BKGD}$ ~ -1.5×10$^{-6}$ emu.g$^{-1}$.Oe$^{-1}$), while the saturation magnetization of the FM polyhedron sample at $T$ = 5 K is $M_S$ ~ 0.19 emu/g, determined by extrapolation of $M$ versus $H$ in the high field region to the axis of $H$ = 0. For the FM polyhedron sample, in addition, a high field linear PM behaviour has been observed with the susceptibility determined as $\chi_{PH}$ = 2.3×10$^{-6}$ emu.g$^{-1}$.Oe$^{-1}$ ($\Delta\chi_{BKGD}$ ~ -1.02×10$^{-6}$ emu.g$^{-1}$.Oe$^{-1}$). The $M(H)$ curves for the FM polyhedra and PM nanospheres in the low field region, within the range of −5 and 5 kOe, are shown in figure 2(b). It reveals a clear open loop with the coercivity, $H_C$ ~237 Oe, for the FM polyhedron sample. For the PM polyhedra, the magnetic susceptibility is determined as, $\chi$ = 1.0×10$^{-5}$ emu.g$^{-1}$.Oe$^{-1}$ (data not shown here). The susceptibility or saturation magnetization determined by the measurement is listed in table. 1. It is interesting that the undoped $Cu_2O$ samples exhibit such a wide spectrum of magnetic properties, from diamagnetic, paramagnertic, to ferromagnetic. This variation is independent of the sample morphology. Instead, it depends on the synthesis process or even on different runs with the same process. For a pure $Cu_2O$ with an ideal lattice structure, the intrinsic magnetic property is diamagnetic since the $d$ shell of $Cu^{+1}$ is full, i.e., the electron configuration of $Cu^{+1}$ is 3d$^{10}$, and neither $Cu^{+1}$ nor $O^{2-}$ is a magnetic ion. In this sense, we believe that the susceptibility of the diamagnetic octahedron, $\chi_{OH}$ = -0.95×10$^{-6}$ emu.g$^{-1}$.Oe$^{-1}$, is close to its intrinsic value since the contribution from the magnetic impurities is smaller by one orders of magnitude as revealed by the impurity analysis later, see table 1.

**Table 1**. Magnetic susceptibility calculated for the contribution of magnetic impurities according to the expression, $\chi = n\mu_{eff}^2/3k_BT$. In the expression, $n$ is for the impurity concentration. It is in units of ppm as detected by the technique of ICP-AES and is converted to units of atom/g-sample for this calculation. The susceptibility is calculated at $T$ = 5 K, in comparison with the measured value at the same temperature. The experimental values of magnetic susceptibility listed in the last column are corrected for the BKGD diamagnetic signal, $\Delta\chi_{BKGD}$, of the capsules. (Note: For Fe, $\mu_{eff}$ is 2 $\mu_B$ ~ 2.2×10$^{-20}$ emu (bulk) or 4 $\mu_B$(atom), for Co, 1.76 $\mu_B$ (bulk) or 3 $\mu_B$(atom), and for Ni, 0.60 $\mu_B$ (bulk) or 2 $\mu_B$(atom).)

| | $n_{Fe}$ (ppm) | $n_{Co}$ (ppm) | $n_{Ni}$ (ppm) | $\chi$(emu/g-Oe)[a] $T$ = 5K (bulk $\mu_{eff}$) | $\chi$(emu/g-Oe)[b] $T$ = 5K (atomic $\mu_{eff}$) | Measurement $T$ = 5 K |
|---|---|---|---|---|---|---|
| Octahedraon | 97.5 | < 2 | < 2 | 2.1×10$^{-7}$ | 6.9×10$^{-7}$ | $\chi_{OH}$=-9.5×10$^{-6}$ |
| Nanosphere | 148.30 | 9.82 | 404.81 | 3.9×10$^{-7}$ | 1.8×10$^{-6}$ | $\chi_{NS}$=2.2×10$^{-5}$ |
| Polyhedron | 649.48 | 886.60 | 25.26 | $M_S$ ~0.292emu/g or $\chi$ ~2.6×10$^{-6}$ | 8.1×10$^{-6}$ | $M_S$ ~0.19 emu/g and $\chi_{PH}$ =2.3×10$^{-6}$ |

[a] Value calculated using the bulk value of magnetic moments for the impurities
[b] Value calculated using the atomic value.

The ZFC and FC $M(T)$ measurements of the PM nanospheres are shown in figure 3(a). For the ZFC measurement, the sample was cooled under zero applied field down to 5 K, then, the data were collected in a small field of 90 Oe as the sample was warmed up. For the FC curve, the data were recorded in the same field during the cooling process. A very thin layer of antiferromagnetic (AFM) surface CuO has been demonstrated to exhibit excessive moment in a previous experiment [34]. The major feature is a peak or maximum showing up in the $M_{ZFC}(T)$ curve at low temperature, $T_F$ ~6.4 K, with an enhanced irreversibility for the $M(H)$ hysteresis loop at $T < T_F$. This is not observed with the $M_{ZFC}(T)$ curve in the present experiment. In addition, the results of HRTEM and XRD do not reveal any evidence for the presence of a surface CuO phase [32]. It indicates that the paramagnetism with the nanosphere sample is unlikely to be attributed to the excessive moments of a thin surface layer of CuO arising from the oxidation of Cu$_2$O. The ZFC and FC $M(T)$ measurements of the FM polyhedra are shown in figure 3(b). The ZFC measurement was carried out by the same procedure as that for the PM nanospheres. For the FC curve, however, the sample was first cooled in a field of $H_{COOL} \sim 20$ kOe down to $T = 5$ K, and then the applied field was reduced to $H_{app} \sim 90$ Oe for the data collection as the sample was warmed up. The $M_{ZFC}(T)$ and $M_{FC}(T)$ curves separate widely from each other with the temperature going up to $T = 380$ K. This is a typical behaviour for FM materials, such as Co [35] and Ni [36]. It indicates that the blocking temperature and, hence, the Curie temperature of the observed ferromagnetism is higher than 380 K. By linear extension of the $M_{ZFC}(T)$ and $M_{FC}(T)$ curves to the regime of higher temperature, the Curie temperature is then estimated as $T_C \sim 455$ K at the intersection, see figure 3(b).

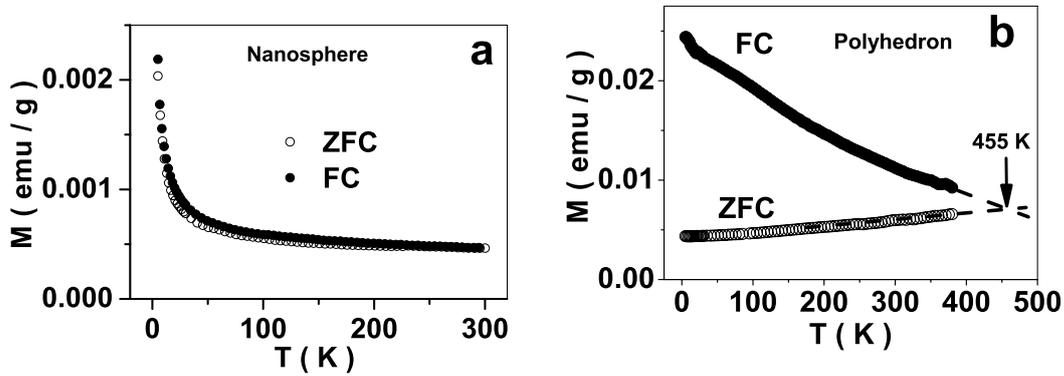

Figure 3. ZFC and FC $M(T)$ curves for the sample of **(a)** PM nanospheres and **(b)** FM polyhedra, measured in the applied field of 90 Oe.

## 4. Characterization of magnetic impurity and analysis

In order to account for the magnetic moments observed with the 4 different powder samples by the magnetization measurements, the magnetic impurity level, in particular, the concentration of Fe, Co and Ni, has been analysed by EDX and ICP-AES. For the EDX analysis, the powdered sample was ultrasonically dispersed in aqueous ethanol, then, several drops of the solution were placed on a quartz glass and dried in air. In order to enhance the electrical conductivity, the quartz glass with the sample was sprayed with Au before observation. The EDX analysis was carried out at five analysis dots randomly selected over the sample. The EDX spectrum for the FM polyhedra is shown in figure 4. The Si and Au peaks are attributed to the quartz glass and the distribution of the sprayed Au. The single peak with energy slightly below 2 keV arises from both Si and Au which overlap each other. With this spectrum, there is no signal for any magnetic impurity. In particular, it does not reveal any sign for the presence of Fe, Co, and Ni, as marked in the spectrum. It indicates that the concentration of any possible magnetic impurity is below a few parts in a thousand.

Further analysis on the impurity concentration has been performed by ICP-AES for the diamagnetic octahedra, PM nanospheres, and FM polyhedra. For the diamagnetic octahedra with $\chi_{OH}$ = -9.5×10$^{-6}$ emu.g$^{-1}$.Oe$^{-1}$, the concentration of Fe is $n_{Fe}$ ~97.5 ppm ~1.05×10$^{18}$ Fe-atom/g-Cu$_2$O ~2.5×10$^{-4}$ Fe-atom/mol-Cu$_2$O, and less than 2 ppm for Co and Ni. At $T$ = 5 K, the contribution to the magnetic susceptibility from the detected Fe impurity is estimated as $\chi = n_{Fe}\mu_{eff}^2/3k_BT$ ~2.1×10$^{-7}$ emu.g$^{-1}$.Oe$^{-1}$ by the assumption that each Fe atom contributes an effective localized moment of $\mu_{eff}$ = 2.2 $\mu_B$ ~2.0×10$^{-20}$ emu for the bulk phase [37], or it is estimated as about $\chi$ ~ 6.9×10$^{-7}$ emu.g$^{-1}$.Oe$^{-1}$ with each Fe atom contributing an effective moment of $\mu_{eff}$ = 4 $\mu_B$ ~3.7×10$^{-20}$ emu for an isolated atom [37]. These values are provided in table 1. Thus, for the diamagnetic octahedra, the contribution to the magnetic susceptibility from the magnetic impurities appears to be a higher order effect, in comparison with the value, $\chi_{OH}$ = -9.5×10$^{-6}$ emu.g$^{-1}$.Oe$^{-1}$. It is therefore reasonable to assume that $\chi_{OH}$ = -9.5×10$^{-6}$ emu.g$^{-1}$.Oe$^{-1}$ is the susceptibility for the background diamagnetism of an impurity-free Cu$_2$O sample, as discussed in the preceding section.

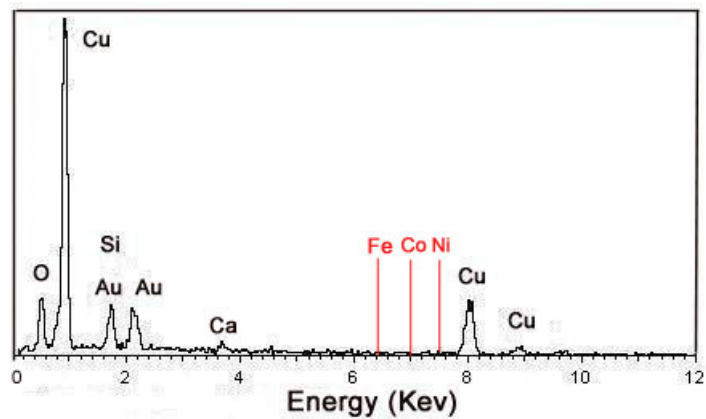

Figure 4. EDX spectrum of the FM polyhedra. No signal is observed at the position marked as Fe, Co, and, Ni.

For the PM nanospheres and the FM polyhedra, the impurity concentrations of Fe, Co, and Ni and the magnitude of the corresponding magnetic susceptibility or saturation magnetization are also analysed and listed in table 1. In all cases, the magnetic impurities cannot account for the magnetism observed in experiment.

Usually, the magnetic property is intrinsic for a substance. It is independent of the sample morphology. However, extrinsic effects, such as the presence of magnetic impurities or crystal defects, are known to alter the magnetic properties. These extrinsic properties depend heavily on the synthesis process and, perhaps, the procedure of measurements due to contamination. In the experiment described above, the magnetic susceptibility for the four $Cu_2O$ samples under investigation is different from sample to sample. It obviously depends on the synthesis process and even on each individual run by the same process. It indicates that the variation of the magnetic properties for the samples is attributed to the extrinsic effects. According to the above analysis, the detected magnetic impurities of Fe, Co, and Ni with concentration on the order of a few hundred ppm or less are not enough to account for the magnetic moments exhibiting the observed susceptibility of the PM sample or the saturation magnetization of the FM sample. Therefore, lattice defects are one of the possible candidates to explain the unaccounted magnetic moments. In the following, we have carried out numerical calculations by DFT to investigate the excessive magnetic moments due to the presence of oxygen and cation vacancies. The dependence on the concentration of defects is also studied.

## 5. Cation vacancy-induced magnetic moments and magnetic ground state by DFT calculations

Numerical calculations have been performed to evaluate the magnetic moment arising from the crystal defects, including the oxygen ($V_O$) and the cation vacancies ($V_{Cu}$). The calculation is by the density functional theory (DFT) within the general gradient approximation (GGA) [38]. To reduce the calculation time, we construct $2 \times 1 \times 1$, $2 \times 2 \times 1$, and $2 \times 2 \times 2$ supercells containing 12, 24, and 48 atoms, respectively, to perform the calculation with different local concentration of vacancies. The concentration of $V_{Cu}$ (or $V_O$) is defined as the number of $Cu^{1+}$ (or $O^{2-}$) vacancies divided by all of the Cu (or O) atoms in a supercell. A procedure of full geometric optimization has been performed by using an ultrasoft pseudopotential [39] with the plane-wave package, CASTEP [40]. A plane-wave cutoff energy of 300 eV and 2 × 2 × 2 Monkhorst-Pack $k$-point mesh are used. The tolerance of the force on each atom is 0.5 eV-nm$^{-1}$. For a pure $Cu_2O$ unit cell, the optimized lattice constant by the calculation is 0.429 nm. This is in good agreement with the value of 0.4267 nm determined by the XRD. The total static energy is calculated with a higher cutoff energy of 380 eV.

The density of state (DOS) has been calculated for the electronic structure of one $V_O$ in a 2 × 2 × 2 supercell. The corresponding local concentration of oxygen vacancies is 6.25%. The presence of a $V_O$ will result in the elongation of the surrounding four Cu-O bonds, from 0.186 to 0.202 nm. However, no net local moment has been found. This indicates that the experimentally observed magnetism is unlikely to arise from the presence of $V_O$. The effect of $V_{Cu}$ has been calculated by the above-mentioned procedure. It is interesting to find that a residual moment of 0.013 $\mu_B$ is obtained corresponding to the cation vacancy concentration, $n_C$ ~3.13%. For the case with larger local concentration of $V_{Cu}$, calculations are carried out by using 2 × 1 × 1 and 2 × 2 × 1 supercells to save the computation time. A larger total magnetic moment is obtained by increasing the vacancy concentration, indicating that the magnitude of the magnetic moment is positively correlated with the $V_{Cu}$ concentration.

The calculated DOS for a perfect $Cu_2O$ supercell and a $Cu_2O$ supercell with $n_C$ ~ 25% are plotted in figures 5(a) and (b), respectively. The valence band is clearly spin-split in the latter case, corresponding to a net effective magnetic moment of about $\mu_{eff}$ = 1.0 $\mu_B$ per vacancy. From the Mulliken population analysis [41], which is a partitioning scheme to allocate the probability of electron distribution among the atoms, inter-atomic bonds and orbitals of a molecular entity, about 67% of the $V_{Cu}$-induced magnetic moments are due to the Cu 3$d$ electrons, and the rest, to the O 2$p$ electrons. The partial DOSs of the Cu 3$d$ and O 2$p$ electrons of $Cu_2O$ with $n_C$

~25% are plotted in figures 5(c) and (d), respectively.

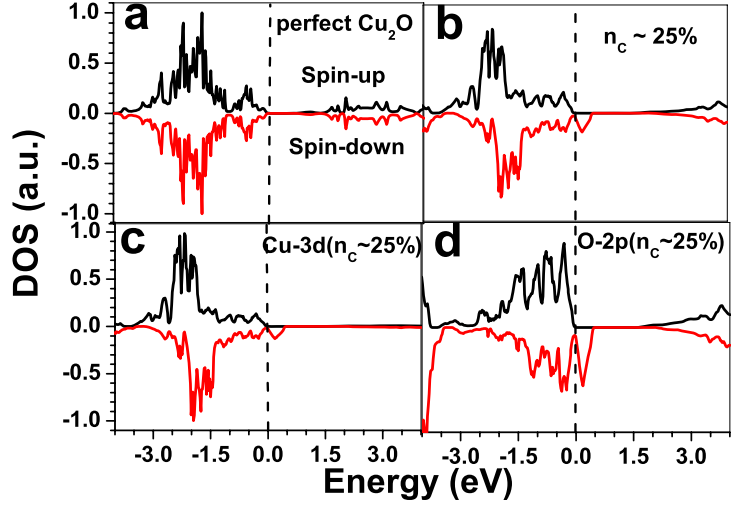

Figure 5. **(a)** Total spin DOS for a perfect $Cu_2O$. **(b)** DOS for the $Cu_2O$ with $n_C$ = 25% . **(c)** Partial spin DOS for the Cu-3$d$ electrons with $n_C$ = 25%. **(d)** Partial spin DOS for the O-2$p$ electrons with $n_C$ = 25%.

The ground state magnetic properties and the local magnetic moments of $Cu_2O$ with various values of $n_C$ have been further calculated. It is found that the magnetic properties depend on the concentration, $n_C$, and the spatial arrangement of the cation vacancies. For a large $n_C$, the FM ordering state is energetically favourable. For instance, with $n_C \sim$ 25%, the total energy of the FM state is lower than that of the AFM state by $\Delta E = E_{FM}-E_{AFM} = -0.070$ eV. It indicates that the FM ordering temperature, $T_C$, is higher than 800 K if the calculated value of $\Delta E$ is taken literally. It is much higher than $T_C \sim$ 450 K for the FM polyhedron sample estimated by the FC and ZFC $M(T)$ measurements shown in figure 3(b). The corresponding magnetic moment is about 0.989 $\mu_B$ per $V_{Cu}$. With the concentration of cation vacancies, $n_C$, and the calculated effective magnetic moment per vacancy, $\mu_{eff}/V_{Cu}$, the saturation magnetization of the FM state is estimated as $M_S = n_C\mu_{eff} \sim$4.83 emu.g$^{-1}$. See table 2 for the values by calculation.

The energy difference between the FM and AFM states is reduced with decreasing $n_C$. With $n_C \sim$12.5%, the energy for the FM state is $\Delta E = E_{FM}-E_{AFM} = -0.014$ eV, which is equal to $T_C \sim 160$ K, for a vacancy separation of 0.430 nm and - 0.002 eV for a larger vacancy separation of 0.603 nm. The magnitude of the coupling energy for the latter is close to zero at $T = 0$ K. Therefore, at a finite temperature, it is expected to exhibit PM properties. The calculation result indicates that the cation-vacancy-induced magnetic moments possibly exhibit PM properties with a separation between defects larger than 0.603 nm, and substantial magnetic interaction is expected at a defect separation smaller than 0.603 nm with the local vacancy concentration of $n_C \sim 12.5\%$. The local magnetic moment also decreases from 1 $\mu_B$ per vacancy for $n_C \sim 25\%$ to 0.709 and 0.777 $\mu_B$ per vacancy corresponding to the two defect separations of 0.430 and 0.603 nm, respectively, as listed in table 2.

| $n_C$ (%) | $E_{FM}-E_{AFM}$ (eV) | Magnetic Coupling | $\mu_{eff}/V_{Cu}$ ($\mu_B$) | $d(V_{Cu}-V_{Cu})$ (nm) | PM $\chi$(emu/g-Oe) | FM $M_S$(emu/g) |
|---|---|---|---|---|---|---|
| 6.25 | 0.005 | AFM | 0.003 | 0.430 | $4.9 \times 10^{-11}$ | |
| 12.5 | -0.014 | FM | 0.709 | 0.430 | $(6.5 \times 10^{-6})^a$ | 1.90 |
| 12.5 | -0.002 | FM | 0.777 | 0.603 | $5.1 \times 10^{-6}$ | |
| 25 | -0.070 | FM | 0.989 | 0.419 | $(2.15 \times 10^{-5})^a$ | 4.83 |

$^a$Value estimated for the total calculated magnetic moments assumed showing PM property.

**Table 2.** Results by DFT calculation for various cation concentrations, $n_C$, for $Cu_2O$, including the energy difference between the FM and AFM states, the magnetic moment per vacancy, and the optimized $V_{Cu}$-$V_{Cu}$ distance in $Cu_2O$.

When $n_C$ is further reduced to 6.25% with a defect separation of 0.430 nm, the ground state reveals a slightly AFM property with $\Delta E = 0.005$ eV. The magnetic state is therefore paramagnetic at $T \sim 50$ K for this defect configuration. It can be expected that with an even smaller $n_C$ and a larger inter-vacancy distance, the system will show a PM behaviour at even lower temperature. The susceptibility estimated for this configuration is very small, $\chi = 4.9\times10^{-11}$ emu.g$^{-1}$.Oe$^{-1}$. It is smaller than the value observed in experiments by several orders of magnitude.

## 6. Discussion

According to the experimental results in the present work, the concentration of magnetic impurities such as Fe, Co and Ni is too small to account for the observed magnetism for the PM octahedra and FM polyhedra. The DFT calculations, on the other hand, show that the cation vacancies in the Cu$_2$O crystal may account for a good part of the observed magnetic moments. In addition, with a local concentration of $n_C \sim$ 25%, $T_C$ of the induced FM state can possibly reach 800 K, much higher than the estimated 455 K in experiment for the FM polyhedra. The itinerant nature of the defect-defficiency-induced electron state and its role to mediate the indirect exchange interaction between localized magnetic impurities of Fe, Co and Ni as described by the RKKY mechanism are not explored in the present work. Additional contribution to the FM coupling might also arise from this process. This may be responsible for the FM coupling as observed with one of the polyhedron samples.

The cation vacancies are one of the possible causes of the unaccounted magnetism of the magnetic impurities according to the DFT calculations. However, the calculations are still too rough, as they are limited by the computing capacity for a large enough size of supercell, to reach any reliable quantitative conclusion, such as the magnitude of the induced magnetic moments per vacancy which depends on the local concentration of the vacancy, the ground state coupling energy for a FM or an AFM ordering state which depends on the distribution configuration of the vacancy, or even the doping effect of the magnetic impurities on the local band structure of the electron distribution, etc. It would be interesting in the future to probe directly the local distribution of the magnetic moments arising from the cation deficiency within the sample to further study the magnetic properties of the undoped Cu$_2$O semiconductor.

## 7. Conclusion

In conclusion, we report both experimental and theoretical investigations on the magnetic properties of undoped $Cu_2O$. For the experiment, four different undoped $Cu_2O$ samples with nanospheres of 200 nm diameter, polyhedra of 800 nm in size and octahedra of 1 μm in size, have exhibited different magnetic properties. The variety in the magnetic property from sample to sample, which does not depend on the morphological forms of the particles, indicates that the observed magnetic properties are not an intrinsic effect. The contribution to the magnetic moments from the magnetic impurities is too small to account for the observed magnetism by about one order of magnitude. According to the *ab initio* DFT calculations, a residual magnetic moment might possibly arise from the presence of cation vacancies in the ideal lattice of diamagnetic $Cu_2O$. In addition, different magnetic ground states are predicted, from PM to FM states, depending on the concentration and spatial distribution of the cation vacancies. An FM property shows up as the local concentration of cation vacancies exceeds 12.5%. Since the magnetic properties of the sample depend heavily on the preparation procedure, as demonstrated by the experiment, the presence of the cation vacancies and the variations of the vacancy concentration in the undoped $Cu_2O$ lattice, which vary easily depending on the synthesis process and conditions, may account for the wide spectrum of magnetic properties observed in the present experiment. More work in the future is needed not only in experiment but also in theory to study the property of the observed magnetism with the undoped $Cu_2O$.

**Acknowledgements**

This work was supported by the NSFC (Grant No. 10874006, 10434010, 90606023, 50725208, and 20731160012), the program for New Century Excellent Talents in University (NCET-04-0164), the State Key Project of Fundamental Research for Nanoscience and Nanotechnology (2006CB0N0300).

References

[1] Pearton S J, Abernathy C R, Overberg M E, Thaler G T, Norton D P, Theodoropoulou N, Hebard A F, Park Y D, Ren F, Kim J and Boatner L A 2003 *J. Appl. Phys.* **93** 1

[2] Kale S N, Ogale S B, Shinde S R, Sahasrabuddhe M, Kulkarni V N, Greene R L and Venkatesan T 2003 *Appl. Phys. Lett.* **82** 2100

[3] Wei M, Braddon N, Zhi D, Midgley P A, Chen S K, Blamire M G and MacManus-Driscoll J L 2005 *Appl. Phys. Lett.* **86** 072514

[4] Liu Y L, Harrington S, Yates K A, Wei M, Blamire M G, MacManus-Driscolla J L and Liu Y C 2005 *Appl. Phys. Lett.* **87** 222108

[5] Zener C 1951 *Phys. Rev* **81** 440

[6] Zener C 1951 *Phys. Rev* **83** 299

[7] Rudermann M A and Kittel C 1954 *Phys. Rev* **96** 99

[8] Kasuya T 1956 *Prog. Theor. Phys.* Japa(Kyoto) **16** 45

[9] Yosida K 1957 *Phys. Rev* **106** 893

[10] Dietl T, Ohno H, Matsukura F, Cibert J and Ferrand D 2000 *Science* **287** 1019

[11] Dietl T, Ohno H and Matsukura F 2000 *Phys. Lett.B* **63** 195205

[12] Dietl T 2007 *J. Phys.: Condens. Matter* **19** 165204

[13] Schlafer H L and Gliemann G *Basic principles of ligand field theory* (John Wiley & Sons Ltd. 1969) P.120

[14] Klemm W and Schüth W 1931 *Z. anorg. allg. Chem.* **203** 104

[15] Perakis N and Serres A 1955 *J. Phys. Radiumn* **16** 387

[16] O'Keeffe M and Stone F S 1962 *Proceedings of the Royal Society of London. Series A, Mathematical and Physical Sciences* **267** 501

[17] Venkatesan M, Fitzgerald C B and Coey J M D 2004 *Nature* **430** 630

[18] Pemmaraju C D and Sanvito S 2005 *Phys. Rev. Lett.* **94** 217205

[19] Elfimov I S, Yunoki S and Sawatzky G A 2002 *Phys. Rev. Lett.* **89** 216403

[20] Yoon S D, Chen Y J, Yang A, Goodrich T L, Zuo X, Arena D A, Ziemer K, Vittoria C and Harris V G 2006 *J. Phys.: Condens. Matter.* **18** L355

[21] Hong N H, Sakai J, Poirot N and Brizé V 2006 *Phys. Rev.* B **73** 132404

[22] Sundaresan A, Bhargavi R, Rangarajan N, Siddesh U and Rao C N R 2006 *Phys. Rev.* B **74** R161306

[23] Osorio-Guillen J, Lany S, Barabash S V and Zunger A 2006 *Phys. Rev. Lett.* **96** 107203

[24] Abraham D W, Frank M M and Guha S 2005 *Appl. Phys. Lett.* **87** 252502

[25] Rao M S R, Kundaliya D C, Ogale S B, Fu L F, Welz S J, Browning N D, Zaitsev V, Varughese B, Cardoso C A, Curtin A, Dhar S, Shinde S R, Venkatesan T, Lofland S E and Schwarz S A 2006 *Appl. Phys. Lett.* **88** 142505

[26] Guo L F and Murphy C J 2003 *Nano Lett.* **3** 231

[27] White b, Yin M, Hall A, Le D, Stolbov S, Rahman T, Turro N and O'Brien S 2006 *Nano Lett.* **6** 2095

[28] Tan Y W, Xue X Y, Peng Q, Zhao H, Wang T H and Li Y D 2007 *Nano Lett.* **7** 3723

[29] Xu H L and Wang W Z 2007 *Angew. Chem. Int. Ed.* 46 1489

[30] Lu C H, Qi L M, Yang J H, Wang X Y, Zhang D Y, Xie J L and Ma J M 2005 *Adv. Mater.* 17 2562

[31] Jiao S H, Xu L F, Jiang K and Xu D S 2006 *Adv. Mater.* **18** 1174

[32] Zhang J T, Liu J F, Peng Q, Wang X and Li Y D 2006 *Chem. Mater.* **18** 867

[33] JCPDS-International Center for Diffraction Data, Card No. 77-0199 (unpublished).

[34] Li Q, Zhang S-W, Zhang Y and Chen C P 2006 *Nanotechnology* **17** 4981

[35] He L, Chen C P, Liang F, Wang N and Guo L 2007 *Phys. Rev. B* **75** 214418

[36] He L, Zheng W Z, Zhou W, Du H L, Chen C P and Guo L 2007 *J. Phys.: Condens. Matter* **19** 036216

[37] Billas I M L, Chatelain A and de Heer W A 1994 *Science* **265** 1682

[38] Perdew J P, Burke K and Ernzerhof M 1996 *Phys. Rev. Lett.* **77** 3865

[39] Vanderbilt D 1990 *Phys. Rev.* B **41** R7892

[40] Milman V, Winkler B, White J A, Pickard C J, Payne M C, Akhmatskaya E V and Nobes R H 2000 *Int. J. Quantum Chem.* **77** 895

[41] Mulliken R S 1955 *J. Chem. Phys.* **23** 1833